# Critical Appraisal of Artificial Intelligence-Mediated Communication in Language Education

Dara Tafazoli

The University of Newcastle, Australia

Dara.Tafazoli@newcastale.edu.au

## Abstract

Over the last two decades, technology use in language learning and teaching has significantly advanced and is now referred to as Computer-Assisted Language Learning (CALL). Recently, the integration of Artificial Intelligence (AI) into CALL has brought about a significant shift in the traditional approach to language education both inside and outside the classroom. In line with this book's scope, I explore the advantages and disadvantages of AI-mediated communication in language education. I begin with a brief review of AI in education. I then introduce the Intelligent CALL (ICALL) and give a critical appraisal of the potential of AI-powered automatic speech recognition (ASR), Machine Translation (MT), Intelligent Tutoring Systems (ITSs), AI-powered chatbots, and Extended Reality (XR). In conclusion, I argue that it is crucial for language teachers to engage in CALL teacher education and professional development to keep up with the ever-evolving technology landscape and improve their teaching effectiveness.

# 1. Introduction

Artificial Intelligence (hereafter, AI) has been progressively utilized in education in recent years, leading to a surge in research and applications of AI in education (henceforth, AIED) (Luckin et al., 2016). Research on AIED is interdisciplinary, involving AI, pedagogy, psychology, and other related disciplines (Luckin et al., 2016; Steenbergen-Hu & Cooper, 2014). The objective of AIED research is to enhance the fields of AI, cognitive science, and education by incorporating computer-supported education (Conati et al., 2002). A number of AIED applications are being applied to adaptive learning and evaluation in order to enhance educational effectiveness and efficiency, modify teaching approaches in real-time, and gain a deeper insight into how students acquire knowledge (Beal et al., 2010; Chassignol et al., 2018; Shute & Psotka, 1996; VanLehn et al., 2007), in various fields like language education.

In this chapter, I concentrated specifically on the integration of AI in language education. AI-based tools applied in language education are part of Computer-Assisted Language Learning (CALL) and Intelligent CALL (ICALL) (Tafazoli et al., 2019). Based on the scope of the book (i.e., Computer-Mediated Communication or CMC), I chose to shift the attention away from the device (e.g., computer) to the positive and negative influences of 'mediated' processes in CMC from a pedagogical perspective to phase out the use of the word 'computer' and reassert that CMC is about the study of mediation and the focus of CMC research should be on human-centered processes, as suggested by Carr (2020).

# 2. Intelligent Computer-Assisted Language Learning (ICALL)

In the realm of language studies, particularly in the field of language education, artificial intelligence has become increasingly prevalent due to the progress made in Natural Language

Processing (abbreviated as NLP), deep learning, and networked learning (Fryer & Carpenter, 2006). With the evolution from CALL to ICALL, there has been a significant enhancement in the quality of interaction between students and computers (Kannan & Munday, 2018). This shift has allowed for a more sophisticated and personalized approach to language education, benefitting both students and educators alike.

The advantages of ICALL arise from the potential of AI to customize digital language learning for each learner (Tafazoli & Gómez-Parra, 2017). ICALL is expected to bring several benefits, including the ability for learners to advance at their preferred speed and to obtain instant feedback, which can serve as a powerful motivator (Huang et al., 2023). This personalization could result in decreased time and cost and reduced frustration for learners. The system can also personalize topic repetition, focusing on topics that learners have found challenging. It provides a swift and objective evaluation of a learner’s progress, as well as a deeper comprehension of their preferences and approaches to learning. By using big data processing algorithms and machine learning algorithms, each learner’s behavior can be dynamically adjusted to maximize these advantages. Furthermore, ICALL has the potential to accurately predict a learner’s future performance and objectively evaluate various teaching tools, including texts, lectures, assignments, and tests. The algorithms have the capacity to evaluate the learner’s aptitudes and limitations and to generate a tailored collection of study materials for every session. In addition, the algorithm can learn from the behavior of both individual and collective learners, strengthening its predictive capabilities (Campbell-Howes, 2019).

According to Schulze (2008), the areas of AI research most pertinent to ICALL are NLP, user modeling, expert systems, and Intelligent Tutoring Systems (ITSs). NLP is concerned with two key aspects of understanding and generating natural language. The former involves designing computers that can receive and comprehend spoken or written natural language input, while the latter aims to develop computers that can generate natural language output, irrespective of whether it is spoken or written (Fryer & Carpenter, 2006). These functions, exemplified by systems like chatbots, entail processing various features of natural language elements, including graphology, phonology, morphology, syntax, semantics, and pragmatics. User modeling aims to tailor computational systems to their users, with the primary objective being to adapt to users' needs. This involves observing user behavior by gathering, retaining, and scrutinizing data from their past task responses. Additionally, user modeling seeks to predict future behavior by tracking personal memory curves. Expert modeling, in addition to user modeling, is a crucial element of intelligent tutoring systems. Statistical and predictive analysis are both involved in user and expert modeling using big data (Fryer & Carpenter, 2006).

ICALL tools have various applications in language education, including but not limited to machine translation (MT), AI-powered virtual language exchange platforms and language learning communities, text-to-speech technology, intelligent content recommendations, chatbots, automated grading systems, adaptive learning games, pronunciation analysis tools, automatic speech recognition (ASR), sentiment analysis tools, speech synthesis technology, intelligent tutoring system (ITS), Extended Reality (XR), intelligent writing assistants, AI-powered interactive textbooks, language learning analytics, cognitive learning, gamified learning platforms, predictive analytics, and AI-powered language learning apps. In the following, I pedagogically reflected on some of the abovementioned ICALL tools based on the literature and provided some recommendations for language teachers.

AI-powered *automatic speech recognition (ASR)* technology can improve students' oral proficiency and fluency in a foreign language (e.g., Chen, 2022; Dai & Wu, 2023; Evers & Chen, 2021, 2022; Mroz, 2018; Song, 2020; van Doremalen et al., 2016). Many of the ASR software can be obtained for free (e.g., Google Speech Recognition, Windows Speech Recognition, Siri assistant, iFlyRec, and AT&T Watson) (Evers & Chen, 2022). These programs have access to a large speech database, which improves their ability to decode speech. Additionally, the free availability of these programs makes them easy to use for classroom or self-study purposes, particularly for constant corrective feedback and self-monitoring. Neuroscience studies have shown that when foreign language learners speak, they tend to monitor their speech production from the perspective of their first language (L1) rather than the target language due to their L1 filtering the sounds in their external monitoring system (Meekings & Scott, 2021). This can make it difficult for them to self-monitor their speech in the target language, and while language teachers can provide feedback, it may not be timely or sufficient for individual learners. To address this issue, ICALL tools such as ASR are becoming more widely used in foreign language learning (Bashori et al., 2021; Evers & Chen, 2021; Mroz, 2018).

Despite their advantages, ASR dictation programs have limitations when it comes to pronunciation instruction. While these resources offer extensive exercises and prompt responses, they do not encompass features related to phonetic descriptions, like clarifying the utilization of the vocal apparatus for specific sounds or the variations between target sounds and the user's native language (Liakin et al., 2014). Learners require more assistance to understand pronunciation, which could be why earlier research observed advancements in speaking abilities, but not in listening capabilities (Liakin et al., 2014). Moreover, despite advancements in ASR technology, recognition accuracy is still lower than that of human evaluation (Loukina et al., 2017), especially when noisy environments are present (Evers & Chen, 2022). The Google Speech Recognition system has an accuracy rate of 93% for free non-native speech (McCrocklin et al., 2019), while other systems like Windows Speech Recognition and Siri are less accurate, with rates of 74% and 69%, respectively (Daniels & Iwago, 2017; McCrocklin et al., 2019). Transcription inaccuracies can cause frustration and demotivation among students, as reported by participants in various studies (Liakin et al., 2017; McCrocklin et al., 2019; Mroz, 2018). Efforts may be undertaken in the near future to address this challenge by helping learners exchange viewpoints about their pronunciation, which may be more accurate than software feedback (Evers & Chen, 2022). Also, it should be noted that most ASR programs were not designed to cater to language learning needs and do not offer any support to modify pronunciation or rectify mistakes. Therefore, some scholars suggest that combining ASR software with scaffolding activities could enhance its effectiveness in language teaching (Evers & Chen, 2022; Mroz, 2018).

*Machine Translation* (henceforth, MT), like Google Translate, can be used to translate text and speech between different languages. MT has gained popularity in the realm of foreign language education as well as in daily use over the past few years because of its convenience, multilingual capabilities, affordability, and immediacy (Alhaisoni & Alhaysony, 2017; Briggs, 2018). According to Ducar and Schocket (2018), a novel translation system, which is released by Google as a neural MT, uses statistical methods to identify the most probable match in the target language from a vast amount of data when translating source texts. Consequently, it has achieved notable improvements in accuracy and comprehensibility compared to its predecessor, a phrase-based statistical MT (SMT) system (Briggs, 2018).

Recent studies have emphasized the benefits of utilizing machine translation (MT) in the field of foreign language education, especially in language writing (Fredholm, 2015; Garcia & Pena, 2011; O'Neill, 2016). MT enables students to write more fluently, communicate more effectively, and concentrate more on content in a second/foreign language with fewer errors (Garcia & Pena, 2011; Shadiev et al., 2019). Furthermore, MT helps learners to minimize errors in vocabulary, grammar, syntax, and spelling (Fredholm, 2019; Lee, 2020; Tsai, 2019), thereby producing higher-quality writing (O'Neill, 2016). In spite of MT's assistance being limited to the sentence level in L2 writing, linguistic errors can severely impact the overall quality of L2 writing, thus demonstrating that MT helps with L2 writing by reducing lexicogrammatical errors (Lee, 2020; Tsai, 2019). Additionally, beyond the linguistic domain, several studies have reported a range of advantages of using MT in the affective and metacognitive domains of foreign language learning (Shadiev et al., 2019).

Despite the potential advantages of utilizing MT, language educators frequently view it as an insufficient or potentially detrimental resource when used in teaching foreign languages for various reasons, including ethical concerns or students' excessive reliance on MT (Lee, 2020; Nguyen et al., 2023). Nonetheless, a considerable number of students already utilize MT for various educational purposes and regard it as a valuable tool for language learning (Alhaisoni & Alhaysony, 2017; Briggs, 2018). While MT can improve writing outcomes by reducing errors, it may not lead to language learning without proper pedagogical design (Lee, 2023). Pedagogical designs should not only focus on effectively using MT in tasks but also on cultivating language learning over a longer period. Teachers' concerns are understandable, but they should also be aware of this gap and consider MT as a language learning aid. In order to make informed decisions with regards to pedagogy, it is essential for teachers to have a thorough understanding of the benefits and limitations of current MT technologies, as well as their potential as a tool for language education. Teachers should also consider the impact of highly accurate MT on language learning, including demotivation and potential academic dishonesty (Murtisari et al., 2019). They need to provide clear guidelines to students on ethical use and prepare for the future of language learning classrooms (Nguyen et al., 2023).

*Intelligent Tutoring Systems (ITSs)* employ AI and machine learning technologies to engage with learners and carry out educational tasks. These systems collect information regarding student reactions, create a model of each student's understanding, awareness, motivation, or sentiment, and deliver customized guidance. ITSs feature interfaces for students to interact with throughout the learning activity, allowing for more detailed student modeling and step-level hints and feedback (Mousavinasab et al., 2021). ITSs utilize AI methodologies to support education by following the principles of cognitive psychology and student learning models (Anderson et al., 1995; Shute & Psotka, 1996; Xu et al., 2019). Researchers in education have devoted their efforts to developing teaching methods that enhance teaching outcomes (Graesser et al., 2005; Kolodner, 2002; Luckin et al., 2016). For instance, Graesser et al. (2005) investigated how pedagogical strategies that embraced constructivist approaches could be integrated into ITS instruction, revealing that learning outcomes were inversely proportional to boredom and directly proportional to a state of flow, and also probed the relationship between emotions and the learning process. These systems are interactive, capturing and analyzing learner performance, selecting corresponding tasks, and presenting appropriate information to the learner (Shute & Zapata-Rivera, 2008). This information provides tailored feedback and creates adaptive instructional input during tutoring sessions (Anderson et al., 1995; Atkinson, 1968; Shute & Psotka, 1996; Xu et al., 2019).

ITSs designed for language education generally consist of various components, such as modeling, forecasting, feedback provision, adaptable lessons and activities, and scaffolding (Hung & Nguyen, 2022; McNamara et al., 2007). The system ensures real-time monitoring of individual students' progress and provides necessary assistance as needed (Graesser et al., 2011). Other computer programs that adapt to learners do not utilize complex learning principles or track cognitive and emotional states like ITSs (Graesser et al., 2011). Scholars also applied various types of ITSs in their studies (Khella & Abu-Naser, 2018; Mayo et al., 2000; Michaud et al., 2000). A tool was developed by Michaud et al. (2000) to enhance the literacy skills of deaf high school and college students who communicate in American Sign Language (ASL) as their primary language. The system evaluates the student's written text for mistakes in grammar and provides a tutorial dialogue to suggest the necessary corrections. The system adapts to the user's knowledge level and learning style and uses both English and the user's native language for tutorial instruction. The study showed how ITS successfully created a flexible, multi-modal, and multi-lingual system that improved the literacy skills of deaf students who use ASL. Mayo et al. (2000) introduced a newly developed ITS that instructs students on the mechanical aspects of English capitalization and punctuation. The mechanism necessitates that students engage in an interactive process where they apply capitalization and punctuation to brief passages of text that are initially written in lowercase. It defines the field using a series of limitations that outline the appropriate punctuation and capitalization formats, and provides responses when students deviate from these limitations. During classroom testing of the ITS, a set of students between the ages of 10 and 11 were involved and the results indicate that the students were successful in learning the 25 rules included in the system. In another paper, Khella and Abu-Naser (2018) outlined the design of a digital ITS aimed at helping students overcome difficulties in learning French.

The system aims to provide a compelling introduction to learning French by presenting the purpose of learning the language and generating related problems for students to solve. It also adjusts to the personal progress of every student in actual time. The system offers explicit assistance and can be flexibly adjusted to the needs of each learner. Based on the mentioned features, ITS, as an exemplar of ICALL, has been found to be almost as effective as teachers (VanLehn, 2011). These intelligent tools can boost the functions of teachers and students (Spector et al., 2014).

*AI-powered chatbots* are designed to interact with users and process natural language inputs. Over half a century ago, ELIZA became the pioneering chatbot (Weizenbaum, 1966). Currently, chatbots have gained immense popularity as a highly effective medium for providing information and addressing frequently asked questions (Smutny & Schreiberova, 2020). Chatbots have been employed in educational environments for various purposes in recent times, including sustaining learners' motivation in scientific studies, supporting first-year students with their college experiences, and aiding educators in managing substantial classroom activities (Carayannopoulos, 2018; Schmulian & Coetzee, 2019).

The potential of chatbots in language teaching has attracted the attention of researchers (Chiu et al., 2023; Fryer et al., 2019; Jia et al., 2012; Xu et al., 2021). Chatbot-supported language learning involves using chatbots to interact with students in the target language for daily practice (Fryer et al., 2019), answering questions (Xu et al., 2021), and conducting assessments (Jia et al., 2012). Chatbots can be a valuable tool in language practice for students. They can help reduce shyness and make the learning experience more comfortable for all involved (Fryer & Carpenter, 2006). Additionally, chatbots can help bridge the gap between learners and instructors in online learning environments, which can reduce the transactional distance and improve the overall experience (Huang et al., 2022). Visual chatbot development platforms, such as Dialogflow and BotStar, allow teachers to create customized chatbots without prior programming experience. These platforms provide a design dashboard that enables teachers to script students' learning experiences and meet their learning objectives (Huang et al., 2022). To learn a new language effectively, it's essential to practice speaking and immerse oneself in language contexts, but many students lack motivation and confidence. Researchers have suggested that chatbot-supported activities can create a more engaging and authentic language environment and improve language learning outcomes (Lu et al., 2006). Language educational chatbots generally possess three main characteristics. Firstly, they are available 24/7 to support students (Garcia Brustenga et al., 2018), allowing them to practice language skills at any time that suits them (Haristiani, 2019). Secondly, chatbots can provide students with a broader range of language information than their peers, who may be at a similar proficiency level, including additional expressions, vocabulary, and questions (Fryer et al., 2019). Thirdly, chatbots can function as tireless assistants and relieve teachers of repetitive tasks such as answering common questions and providing continuous language practice (Fryer et al., 2019; Kim, 2018). Chatbots are always available to help students practice speaking and learn the new language.

Although chatbots have proven to be advantageous in language education by decreasing students' anxiety (Ayedoun et al., 2019) and enhancing their participation in language learning (Ruan et al., 2019), the temporary nature of learners' engagement and performance improvement may be due to the novelty effect associated with chatbots (Ayedoun et al., 2019; Fryer et al., 2019). The novelty effect refers to the initial excitement of a new technology that wears off as students become more accustomed to it. Additionally, concerns have been raised about chatbots' limited capabilities. While AI has advanced significantly, designing intelligent dialogue in chatbots remains a challenge for developers (Brandtzaeg & Følstad, 2018). Even small mistakes in student input can lead to irrelevant responses from the chatbot, which may not be able to understand multiple sentences at once as humans can (Kim et al., 2019). This can restrict students' interaction to a pre-set knowledge base (Grudin & Jacques, 2019) and may result in chatbots providing unrelated answers (Haristiani, 2019; Lu et al., 2006).

Another challenge deals with cognitive load limitations (Huang et al., 2022). Cognitive load limitations refer to the additional attention or mental effort which are necessary to complete a task during the learning process (Sweller, 1988). The amount of cognitive burden that students must carry depends on the instructional design of activities supported by chatbots. If the cognitive load is too high, it can interfere with learning outcomes, particularly for low-proficient students (Kim, 2016). Therefore, the use of chatbots must be carefully designed to avoid imposing an excessive cognitive load (Fryer et al., 2019). Teachers should take a leadership role in determining the best way to use chatbots to achieve learning outcomes and mitigate their limitations (Huang et al., 2022). It is imperative to complete a task that involves learning, as it is a crucial part of the learning process. For example, chatbots may be more appropriate for advanced learners, and restricted chatbots can be used to correct spelling errors or check factual knowledge for beginners. Teachers have the ability to establish guidelines for interactions with chatbots, which can assist learners in comprehending the capabilities and limitations of these conversational agents. To address the novelty effect, students can be prepared through a workshop before the first lesson, and multimedia principles and human-like gestures can be employed to enhance students' cognitive processing. Quick buttons can also be used to enhance interactivity and engagement between chatbot and students. These measures can help make the chatbot experience more enjoyable and effective for language learners. Taking into account the present level of technological progress is equally significant when implementing chatbots in language learning.

*Extended Reality (XR)*, including Virtual Reality (VR), Augmented Reality (AR), and Mixed or Merged Reality (MR), can be used to create immersive language learning experiences, allowing students to practice real-world language skills in a simulated environment. Over the last ten years, XR has gained significant popularity. As XR aims to provide realistic simulations, authenticity, a strong sense of presence, and exposure, it has been identified as an essential tool for language learning by researchers in language education (Tafazoli, in press). Several studies have been conducted worldwide to explore the potential benefits of XR in language education (See, Bonner & Reinders, 2018; Godwin-Jones, 2016; Lan, 2020; Peterson & Jabbari, 2022).

CALL researchers have proposed that XR provides language learners with a distinct and innovative learning environment due to its CMC exclusive learning environments (Peixoto et al., 2021). The advantages of XR include enhancing learners' interest, motivation, engagement, and spatial memory and knowledge (Lege & Bonner, 2020; Xie et al., 2021), providing an inaccessible environment, distance learning, and empathy training (Bonner & Reinders, 2018; Lege & Bonner, 2020), reducing distractions (Bonner & Reinders, 2018), linking classroom concepts to the real world (Reinders & Wattana, 2018), facilitating interactions (Bonner & Reinders, 2018), providing a culturally rich and dynamic context (Godwin-Jones, 2016; Yeh & Kessler, 2015), and promoting learners to participate in the construction of their learning environment (Bonner & Reinders, 2018). These are just some of the benefits that XR offers in language education, as suggested by scholars.

Although virtual reality has been shown to have positive effects on language learning, language educators have mixed opinions on its use. Some of the main barriers to the integration of XR in language education are the high cost of VR tools and the need for advanced digital literacy skills (Parmaxi et al., 2017). Lack of VR-specific pedagogy, cognitive demands, and potential immersion-breaking also pose challenges (Lege & Bonner, 2020). To effectively integrate VR into language education, teachers need to introduce it to the classroom before implementing it, as explained by Southgate et al. (2018). Gender should also be considered as an influential variable in VR integration (Southgate et al., 2019). In addition, empirical studies on the merits of XR in teacher education and the design and development of such tools are scarce (Tafazoli, in press). Therefore, it would not be possible to judge the affordability of XR from the teachers and material developers' perspectives. Furthermore, the lack of XR-specific pedagogy which spells out 'why' and 'how' language education stakeholders should constructively and compellingly integrate technology, is crystal clear. In other words, implementing XR without sufficient and efficient teacher training is useless.

In conclusion, the use of ICALL is widely regarded as a vital component for the development of education (Gu et al., 2021; Luckin et al., 2016). However, it is important to recognize that language teachers need to be equipped with the necessary skills and knowledge to effectively integrate these intelligent tools into their teaching practices. This involves providing adequate training and education to ensure that teachers are proficient in this area, and can effectively implement these technologies in their classrooms. The next section delves further into these considerations.

## 4. ICALL and Teacher Education

While human teachers and social interactions outside of the digital realm remain crucial for achieving fluency in language education, the integration of ICALL in language education has led to a redefinition of the roles of teachers and learners (Lam & Lawrence, 2002). As previously reviewed in the same chapter, AI-based systems offer language learners an environment where they have the freedom to choose their own learning path and pace, granting them greater control over the learning process. These systems facilitate the development of learners' decision-making skills, resulting in greater learning autonomy. Through ICALL tools connections with native speakers are easier, and foreign- and second-language learners can intensify their learning without the need for a teacher's direct involvement. This enables learners to become more active participants in the learning process rather than being passive recipients of knowledge.

The integration of AI into language learning systems has revolutionized the teaching approach, enabling a more personalized learning experience. The learner now has the autonomy to make independent decisions and take charge of their progress. This shift in dynamic allows the teacher to take on a new role as a facilitator and supporter of the learner's unique path toward proficiency. As a result, the teacher is no longer the sole authority and decision-maker, but rather works alongside the learner to guide them towards success. This approach has been widely recognized in the academic community, as it brings about a more inclusive and effective learning environment (Bancheri, 2006; Rilling et al., 2005).

Currently, there is a lack of research on the newly emerged AI-powered tools in language education, such as ChatGPT. To the best of my knowledge, no empirical studies have been conducted on the pedagogical impact of large language models in foreign language classes, nor on the attitudes of learners or teachers towards their use. Additionally, there has been no research on teacher training or preparation for integrating AI-powered tools into language classes. It is not necessary to explore the topic of using AI-powered tools in language education completely in isolation or from the beginning. The preparation of teachers for ICALL is a part of CALL teacher education and professional development which has been addressed in multiple book publications (See, Hubbard & Levy, 2006; Son, 2018; Tafazoli & Picard, 2023; Torsani, 2016) and research articles (e.g., Hubbard, 2008, 2018, 2023; Kessler, 2007, 2010; Levy, 1997; Lord & Lomicka, 2011). The general goal of CALL teacher education is to provide language teachers, both present and future, with the necessary technical and pedagogical knowledge and skills to effectively incorporate technology into their classes (Hubbard, 2008; Tafazoli, 2021).

There have been numerous research studies conducted that have revealed a positive attitude among language instructors regarding the integration of CALL and other contemporary technologies in their classrooms. However, despite this positive outlook, many instructors tend to be hesitant in utilizing these technologies to a great extent. Several external factors such as a lack of equipment, technical support, inflexible curriculum, and time constraints can contribute to this reluctance. Additionally, there are internal factors such as a lack of CALL literacy, limited experience with technology as a learner, lack of motivation, difficulty integrating technology with existing teaching practices and learning styles, fear of being outside of their comfort zone, and the fear of losing control over the classroom and students' respect that can also influence this reluctance (See, Tafazoli & McCallum, in press).

Park and Son (2009) discovered in their study that while teachers acknowledged that CALL makes language learning more engaging, they did not believe that they needed to be experts in using computers. Abdelhalim (2016) found that even when teachers integrated technology into their teaching, they mainly used basic applications such as email or web browsing. Therefore, CALL teacher trainers should consider these factors when developing their training programs. Although it may be premature to identify specific ICALL teacher skills or propose ICALL teacher training models, such models will likely emerge in the near future. It will be essential to approach this task realistically and practically. Language teachers do not need to have programming skills or expertise in artificial intelligence to use chatbots or incorporate ICALL practice into their classes.

Several scholars have developed detailed inventories and intricate diagrams of essential abilities for teachers (See, Mishra & Koehler, 2006), which can impose impractical demands on teachers of foreign languages. These materials overlook the fact that such teachers are primarily language educators and professionals. To effectively overcome the aforementioned barriers to successful CALL implementation, adequate and ongoing professional training may be the best solution. Teachers must believe that technology can assist them in achieving educational objectives more efficiently and effectively without disrupting other aspects of classroom management. They must also possess sufficient CALL skills and have unrestricted access to technology.

## 5. Conclusion

Incorporating AI into language education has given rise to the concept of ICALL, which offers a new level of quality in language teaching and learning. AI-based tools can provide a sophisticated educational environment that is more personalized and flexible for learners and teachers. These tools can assist learners in acquiring the knowledge and skills that modern society demands. There are individuals who hold a pessimistic perspective towards the incorporation of AI, fearing that it may obtain complete dominance and transform into an oppressive mentor that directs the content, timing, and manner in which students acquire knowledge, using information gathered without their approval. In contrast, others have a positive view, envisioning learners who control their personal AI tools, which aid them (and their teachers) in better understanding their progress and organizing learning activities (Fryer & Carpenter, 2006).

The eminence of CALL teacher education and professional development should be considered in this situation. Language teachers required to pick up new skills to integrate ICALL tools into their teaching processes effectively and avoid unnecessary workloads and repetitive tasks. The use of tools such as writing assistants and correction systems can support learners. However, it remains to be seen how well-informed language teachers are about ICALL advancements and how frequently they incorporate AI tools into their teaching. Research needs to answer various questions, such as what the preferred AI tools among language teachers are, how they perceive ICALL, and what motivates them to use it. Additionally, identifying the key skills required for AI-enhanced teaching environments and developing appropriate teacher training programs are crucial.